\documentclass[aps,twocolumn,prl,amsmath,amssymb,notitlepage,superscriptaddress,longbibliography,nofootinbib,longbibliography]{revtex4-2}

\usepackage{microtype}
\usepackage[utf8]{inputenc}
\usepackage{physics}                            
\usepackage{bm}									
\usepackage{enumerate}							
\usepackage{amsthm}								
\usepackage{comment}							
\usepackage[caption=false]{subfig}
\usepackage{amssymb}							
\usepackage{t1enc}								
\usepackage{mathtools}							
\usepackage{tensor}								
\usepackage{wrapfig}							
\usepackage{listings}							
\usepackage[usenames,dvipsnames]{xcolor}								
\usepackage[pdftex,colorlinks=true,pdfstartview=FitV,linkcolor= linkcolor,citecolor= linkcolor,urlcolor= linkcolor,hyperindex=true,hyperfigures=false]{hyperref}
\definecolor{linkcolor}{rgb}{0,0,0.6}
\hypersetup{colorlinks=true}
\usepackage[capitalize]{cleveref}               
\usepackage{graphicx,color,xcolor}
\usepackage{amsfonts,amsmath}
\usepackage{xstring}                            
\usepackage{tikz}

\newcommand{\ccite}[1]{\IfSubStr{#1}{,}{Refs.~}{Ref.~}\cite{#1}}

\newcommand{\bfr}{\mathbf{r}}
\newcommand{\bfu}{\mathbf{u}}

\newcommand{\cD}{\mathcal{D}}
\newcommand{\cL}{\mathcal{L}}

\newcommand{\cS}{\mathcal{S}}

\newcommand{\rmd}{\mathrm{d}}

\newcommand{\bfR}{\boldsymbol{R}}

\newcommand{\plm}[1]{{\color{blue}[#1]}}

\begin{document}
\title{Unifying Hydrodynamic Theory for Motility-Regulated Active Matter: \\From Single Particles to Interacting Polymers}
\author{Alberto Dinelli}
\altaffiliation[Corresponding authors.]{ Email: {alberto.dinelli@unige.ch and pierluigi.muzzeddu@unige.ch}}
\affiliation{Department of Biochemistry, University of Geneva, 1211 Geneva, Switzerland}
\affiliation{Department of Theoretical Physics, University of Geneva, 1211 Geneva, Switzerland}
\author{Pietro Luigi Muzzeddu}
\altaffiliation[Corresponding authors.]{ Email: {alberto.dinelli@unige.ch and pierluigi.muzzeddu@unige.ch}}
\affiliation{Department of Biochemistry, University of Geneva, 1211 Geneva, Switzerland}
\affiliation{Department of Theoretical Physics, University of Geneva, 1211 Geneva, Switzerland}

\begin{abstract}
    Understanding how microscopic motility shapes emergent collective behaviors is a challenging task in active matter, especially when self-propulsion is regulated by external cues or via quorum-sensing interactions. 
    To address this problem, we derive a closed hydrodynamics for scalar active matter with spatially-regulated motility, under general hypotheses for the microscopic dynamics of the particles' orientations.
    We show that, at large scales, the contribution of the latter is entirely captured by the autocorrelation tensor of the orientations.
    This allows us to establish a macroscopic equivalence within a broad class of motility-regulated active systems, from single particles to active polymers. 
    %
    %
    Our formalism allows us to reveal a new form of motility-induced phase separation for quorum-sensing active polymers, which we term \emph{anti}-MIPS, where dense phases exhibit \emph{enhanced} activity relative to dilute regions. 
    Our theory shows that anti-MIPS generically arises for motility-regulated agents with internal structure, uncovering the existence of several distinct transition pathways. 
\end{abstract}

\maketitle

In the living world, organisms at all scales adapt their motion in response to environmental cues:
\textit{E. Coli} bacteria adjust their tumbling rate in the presence of chemical gradients~\cite{budrene1991complex,schnitzer1993theory,berg2004coli}, phototactic algae like \textit{C. reinhardtii} reorient their motion under illumination~\cite{polin2009chlamydomonas},  and pedestrians adapt their walking speed with crowd density. 
Beyond responding to external signals, active agents can also produce the cues that regulate their own motion: in quorum-sensing (QS) bacteria, for instance, the secretion and detection of signaling molecules couples motility to the local population density~\cite{miller2001quorum,daniels2004quorum}.

In general, the feedback between activity and external cues, or motility regulation, allows for the emergence of self-organized collective phases in active systems. 
These include motility-induced phase separation (MIPS)~\cite{tailleur2008statistical,cates2015motility,solon2018generalized,zhao2023chemotactic}, chemotactic collapse~\cite{keller1970initiation,chavanis2007critical,pohl2014dynamic,saha2014clusters,o2020lamellar} and dynamic phases fuelled by non-reciprocity~\cite{you2020nonreciprocity,saha2020scalar,dinelli2023nonreciprocity,ouazan2023self,duan2023dynamical,pisegna2024emergent}. 
In addition to biological examples, motility regulation can be exploited in synthetic systems to control the motion of auto-phoretic colloids~\cite{palacci2013living,bauerle2018self,lavergne2019group,fernandez2020feedback}, or to induce pattern formation in engineered bacterial strains~\cite{liu2011sequential,arlt2018painting,frangipane2018dynamic,curatolo2020cooperative}.

To understand the emergent physics of motility-regulated systems, a variety of coarse-graining methods have been developed, relating microscopic dynamics to macroscopic behavior~\cite{fox1986uniform,fox1986functional,schnitzer1993theory,tailleur2008statistical,cates2013when,wittmann2017effective,solon2018generalized,o2020lamellar,duan2023dynamical,dinelli2024fluctuating,burekovic2026active}. 
These approaches typically start from specific models in which the propulsion direction (orientation) follows prescribed dynamics, leading to case-by-case coarse-grained theories. 
Standard examples include run-and-tumble (RTP)~\cite{schnitzer1993theory,berg2004coli,kurzthaler2024characterization}, active Brownian (ABP)~\cite{golestanian2007designing,jiang2010active,theurkauff2012dynamic,palacci2013living}, and active Ornstein–Uhlenbeck (AOUP)~\cite{szamel2014self,martin2021statistical} processes.
As a consequence, no general principle currently identifies which features of the orientational dynamics control the macroscopic theory of these systems.

The situation becomes even more complex for self-propelled agents with internal structure, as active polymers~\cite{isele2015self,ghosh2014dynamics,lin2014dynamics,winkler2017active,bianco2018globulelike,abaurrea2018collective,winkler2020physics,pfreundt2023controlled,dedenon2026importance}. 
There, both orientational dynamics and polymer structure affect the emergent collective behavior, making the connection between microscopic and macroscopic physics unclear beyond specific models~\cite{vuijk2021chemotaxis,muzzeddu2022active,muzzeddu2023taxis,muzzeddu2024migration,valecha2025active}.

\begin{center}
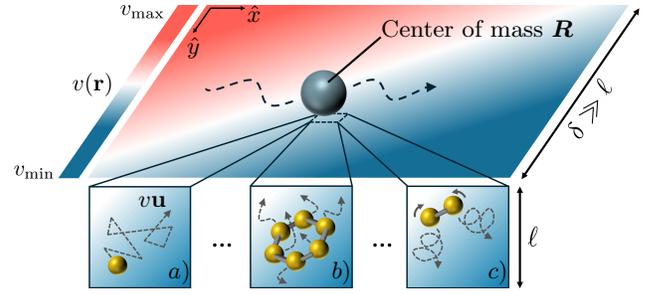
\begin{figure}
\begin{tikzpicture}
    \path (0,-0.07) node [inner sep=0, text width=\columnwidth]{\includegraphics[width=\columnwidth,clip,trim={0 87 0 20}]{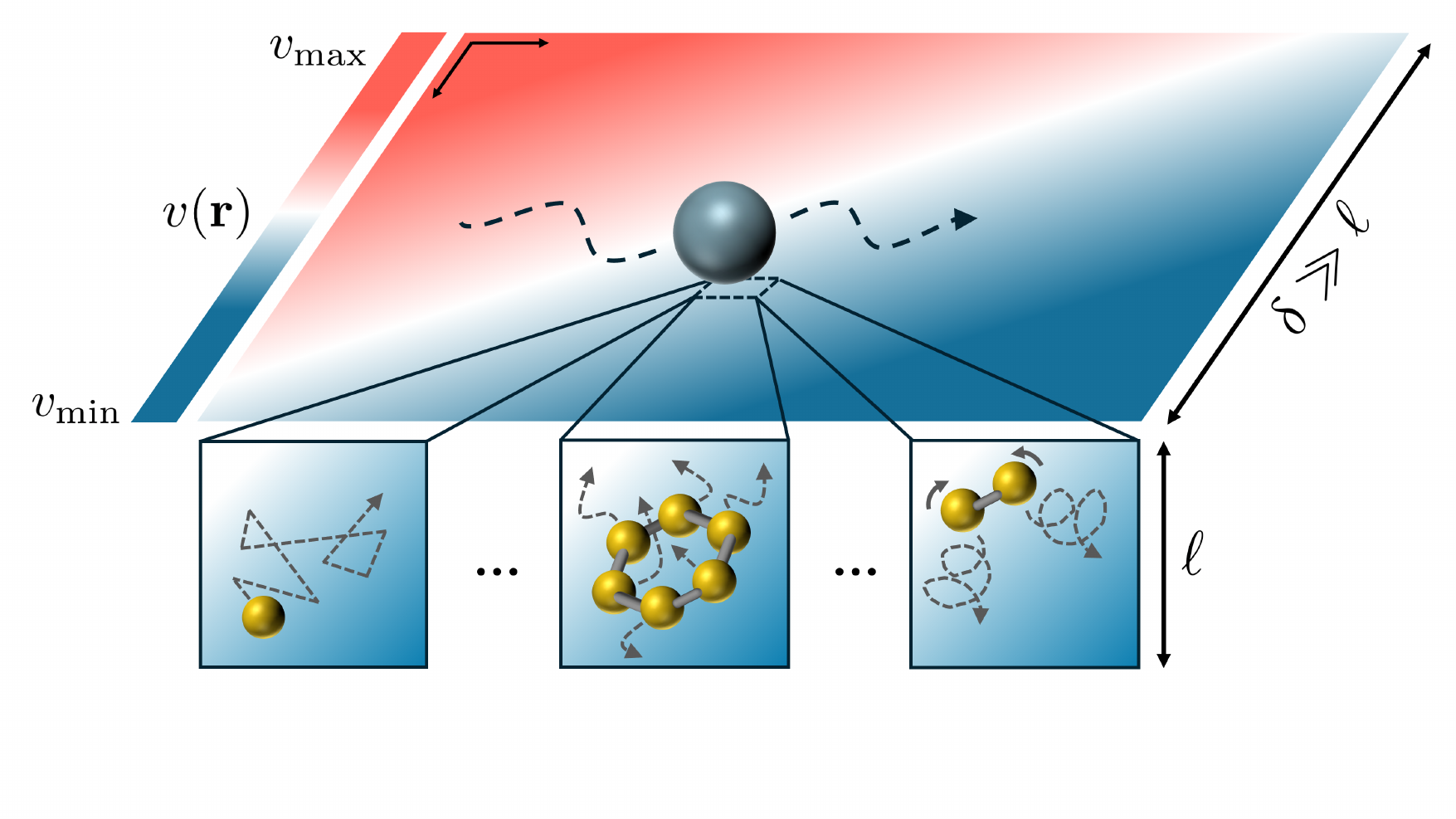}};
    \path (-2.0,-1.75) node [inner sep=0, text width=5pt] {\textcolor{black}{$a)$}};
    \path (-2.4,-0.8) node [inner sep=0, text width=5pt] {\small{\textcolor{black}{$v \bfu$}}};
    \path (0.17,-1.75) node [inner sep=0, text width=5pt] {\textcolor{black}{$b)$}};
    \path (2.25,-1.75) node [inner sep=0, text width=5pt] {\textcolor{black}{$c)$}};
    \path (2.5,1.5) node (CM) [inner sep=0, text width=100pt] {Center of mass $\bfR$};
    \draw [-, thick] (CM.south west)--++(-0.7,-0.6);
    \path (-1.75, 1.25) node [inner sep=0, text width=5pt] {$\hat{y}$};
    \path (-0.95, 1.7) node [inner sep=0, text width=5pt] {$\hat{x}$};
\end{tikzpicture}
    \caption{Emergent large-scale equivalence across active systems with space-dependent self-propulsion speed $v(\bfr)$. At the bottom of our sketch, we show distinct examples of microscopic processes, where each particle propels along an orientation $\bfu$ with speed $v(\bfr)$: (a) single run-and-tumble particle, (b) ring polymer, (c) chiral dimer. At the macroscopic scale $\delta$ over which $v(\bfr)$ varies, there exists a unified Langevin description for the slow mode of the centre of mass, Eqs.~\eqref{eq:FokkerPlanck}--\eqref{eq:diffusion_final}.} 
    \label{fig:sketch}
\end{figure}
\end{center}

\vspace{-0.85cm}

To bridge this gap, in this \textit{Letter} we derive a unified dynamics for a broad class of motility-regulated active systems, ranging from single particles to active polymers, see Fig.~\ref{fig:sketch}. 
Crucially, our framework does not rely on a specific microscopic dynamics for the orientations.
By first considering non-interacting polymers with space-dependent activity, we show that both macroscopic diffusion and drift are solely determined by the auto-correlation tensor of the orientations. 
Building on this result, we derive the hydrodynamics of systems with density-dependent interactions. 
This allows us to reveal a new form of phase separation for active polymers, which we refer to as \emph{anti}-MIPS: in contrast to conventional MIPS, phase coexistence is driven by an \emph{enhancement} of motility at high density, leading to a dense phase that is also the most motile.

Overall, we establish a unified hydrodynamic description across distinct microscopic models, which can be used to design new collective behaviors. 
Details on our derivations and generalizations of the systems discussed here can be found in the companion paper~\cite{dinelli2026PRE}.

\smallskip
\noindent\emph{Active polymer in activity landscape.}---
We consider a single active polymer composed of $N$ interacting particles in $d$ spatial 
dimensions.
Each monomer $i$ is described by a position $\bfr_i$ and an orientation vector $\bfu_i$.
For brevity, we denote by $\Theta = \{\bfu_i\}_{i=0}^{N-1}$ the set of all orientational degrees of freedom.
The polymer topology is described by the connectivity 
matrix $M_{ij} = \deg[i] \delta_{ij} - A_{ij}$, with $A_{ij}$ the adjacency matrix and $\deg[i]$ the number 
of connections of particle $i$.
Interactions between monomers are captured by the quadratic Hamiltonian $\mathcal{H} = \frac{\kappa}{2} \sum_{ij} M_{ij} r_i^\alpha r^\alpha_j$ with elastic coupling $\kappa$, 
where Greek letters denote spatial components and repeated superscripts imply Einstein summation. 
The stochastic evolution of particle $i$ is governed by the overdamped It\^o-Langevin dynamics:
\begin{equation}
    \dot{r}^\alpha_i = - \gamma^{-1} \partial^\alpha_{\bfr_i} \mathcal{H} + v(\bfr_i) u^\alpha_i + \sqrt{2 D_{\rm t}}\xi^\alpha_i\;,
    \label{eq:dynamics-position}
\end{equation}
where $\gamma$ denotes the friction coefficient and $\{\xi^\alpha_i\}$ are independent zero-mean Gaussian white noises with correlations $\langle \xi^\alpha_i(t)\xi_j^\beta(s) \rangle=\delta_{ij}\delta^{\alpha \beta}\delta(t-s)$, accounting for thermal fluctuations.
The noise amplitude in Eq.~\eqref{eq:dynamics-position} is proportional to the thermal diffusivity $D_{\rm t}$.
Each particle $i$ is subjected to a self-propulsion force directed along its orientation $\bfu_i$. 
Motility regulation is modeled as a space-dependent self-propulsion speed $v(\bfr)$~\cite{frangipane2018dynamic,arlt2018painting}.

%
For the purpose of deriving a coarse-grained theory, we find it advantageous to describe the polymer configuration in terms of its center of mass (CM), i.e., $R^{\alpha} := N^{-1} \sum_{i} r_i^\alpha$, and of its normal (Rouse) modes $\chi := \{ \chi^\alpha_i \}_{i=1}^{N-1}$~\cite{doi1988theory}. 
The latter are defined as $\chi^\alpha_i = \sum_{j=0}^{N-1} \varphi_{ij} r_j^\alpha$,
where $\varphi_{ij}$ is the orthonormal matrix that diagonalizes the connectivity $M_{ij}$.
The dynamics of $R^\alpha$ and $\chi$ respectively read:
\begin{align}
    \dot{R}^\alpha &= \frac{1}{N}\sum_{i=0}^{N-1} v(\bfr_i) u^\alpha_i + \sqrt{\frac{2 D_{\rm t}}{N}}\eta_0^\alpha \;,
    \label{eq:R_micro_dyn}\\
    \dot \chi_i^\alpha &= - \lambda_i \chi_i^\alpha + \sum_{j=0}^{N-1} \varphi_{ij} v(\bfr_j) u_j^\alpha + \sqrt{2 D_{\rm t}} \eta^\alpha_i \;,
    \label{eq:relaxationRouse}
\end{align}
where we introduced the relaxation rates $\lambda_i = (\gamma/\kappa) \sigma_i$, with $\{\sigma_i\}$ the eigenvalues of $M_{ij}$.
Moreover, we denoted by $\{\eta_i^\alpha\}$ a set of Gaussian white noises with the same statistical properties as $\{\xi_i^\alpha\}$.
Importantly, Eq.~\eqref{eq:relaxationRouse} identifies a fast relaxation timescale for the Rouse modes: $\tau_\chi \sim \kappa/\gamma$.
The typical size of the polymer is given by the gyration radius $R_g$, where $R_g^2 = N^{-1} \sum_{i=0}^{N-1} \langle |\bfr_i - \bfR|^2 \rangle$, and is directly related to the power spectrum of $\chi$~\cite{doi1988theory}. 

We now turn to the dynamics of the orientations $\Theta$.
As we show below, the large-scale dynamics of the system can be uniquely determined under general conditions on the orientational processes.
For this reason, we remain agnostic on the specific details of the $\Theta$-dynamics, and only assume that it is governed by a Markov process with generator $\mathcal{L}_\Theta$, independent of space~\cite{gardiner2004handbook,pavliotis2008multiscale}. 
We require such process to be ergodic, namely, to admit a single steady-state probability distribution $\psi(\Theta)$
which solves $\cL^\dagger_\Theta \psi = 0$, and impose the average value of the orientations at steady state to be zero, i.e., $\langle u_i^\alpha \rangle_\psi = 0$.

Finally, we introduce the stationary auto-correlation function of the
orientational degrees of freedom, i.e., $\mathbb{C}_{ij}^{\alpha\beta}(t) = \langle u_i^\alpha(t) u_j^\beta(0) \rangle$, where the bracket notation $\langle \cdot\rangle$ denotes the average over the 
$\Theta$-dynamics at steady state. 
To ensure that the angular process has a finite decorrelation time, we require the integrals $\tau^{\alpha\beta}_{ij} := d \int_0^{\infty} \mathbb{C}_{ij}^{\alpha\beta}(t) \rmd t$, 
to be finite for all $\alpha,\beta, i,j$.
The modulus of $\tau^{\alpha\beta}_{ij}$ represents the effective time over which the 
orientations of monomers $i,j$ are correlated along the spatial directions $\alpha,\beta$, while its sign indicates positive or negative correlation. 
We define the persistence time as $\tau_{\rm p} := \max |\tau_{ij}^{\alpha\beta}|$, and introduce its associated persistence length $\ell_{\rm p} = v_0 \tau_{\rm p}$, with $v_0$ the typical scale of the activity $v(\bfr)$. 

\emph{Coarse-graining via multi-scale expansion.}---
Our goal is to derive a closed diffusion-drift dynamics describing the system at the macroscopic scale. 
To do so, we identify $R^\alpha$ as the only slow diffusive degree of freedom, and marginalize the fast orientational and conformational modes $\{\Theta,\chi\}$. 
The dynamics of the latter is characterized by a microscopic lengthscale $\ell \sim \ell_{\rm p} \sim R_g$ and timescale $\tau \sim \tau_{\rm p} \sim \tau_{\chi}$.
A spatio-temporal scale separation then naturally emerges once we assume that the activity landscape $v(\bfr)$ 
varies over a large lengthscale $\delta \gg \ell$, as sketched in Fig.~\ref{fig:sketch}. 
The CM then evolves as a slow diffusive mode over a timescale $\mathcal{T} \sim \delta^{1/2}$ that is much larger than the microscopic time $\tau$.

In light of this scale separation, we introduce the small parameter $\varepsilon := \ell/\delta = \sqrt{\tau/\mathcal{T}}$, which we use to rescale the equation of motion of the active polymer.
As detailed in our companion paper~\cite{dinelli2026PRE}, we then perturbatively expand the stochastic dynamics in Eqs.~\eqref{eq:R_micro_dyn} and ~\eqref{eq:relaxationRouse}.
Our procedure closely follows the homogenization method of Ref.~\cite{pavliotis2008multiscale} and relies on a multiscale expansion of the associated backward Kolmogorov equation.
%
To lowest order in gradients, we obtain a Fokker-Planck equation for the probability distribution $p(\bfR,t)$ of observing the center of mass at position $\bfR$ at time $t$, i.e.,
\begin{equation}
    \partial_{t} p = -\partial^\alpha J^\alpha\;, \quad J^\alpha = V^\alpha p -\partial^\beta(\cD^{\alpha \beta}p)\;.
    \label{eq:FokkerPlanck}
\end{equation}
Here, the drift term $V^\alpha$ and the diffusion tensor $\cD^{\alpha\beta}$ are, respectively, given by~\cite{dinelli2026PRE}:
\begin{align}
\label{eq:drift_final}
    V^{\alpha} &= \frac{\partial^\beta v^2(\bfR)}{2N}\sum_{ijk}  \varphi_{ji} \varphi_{jk}  \int_0^\infty \rmd t\,  e^{-\lambda_j t} \mathbb{C}_{ik}^{\alpha \beta}(t)\;, \\
    \cD^{\alpha\beta}  &= \frac{D_{\rm t}}{N} \delta^{\alpha\beta} + \frac{v^2(\bfR)}{N^2}\sum_{ij}   \int_0^\infty \rmd t\,  \mathbb{C}_{ij}^{\alpha \beta}(t)  \;. 
\label{eq:diffusion_final}
\end{align}
Equations~\eqref{eq:drift_final},~\eqref{eq:diffusion_final} 
relate the macroscopic drift and the diffusion tensor to the microscopic
parameters of the system. 
In particular, the structure of the polymer enters via the Rouse eigenvectors $\varphi_{ij}$ and the relaxation rates $\{\lambda_j\}$.
Note that Eq.~\eqref{eq:diffusion_final} associates the auto-correlations of the active velocity to the macroscopic diffusion via a generalized Green-Kubo relation~\cite{green1954markoff,pavliotis2008multiscale,sharma2016communication,dal2019linear,hargus2021odd}.
Importantly, the orientational auto-correlation tensor
$\mathbb{C}^{\alpha\beta}_{ij}$ is the only relevant feature of the $\Theta$-dynamics affecting the macroscopic description, thus revealing a large-scale equivalence between distinct modes of propulsion.

\smallskip
\noindent\emph{Single particle: effective equilibrium and violations thereof.}---To show the implications of this large-scale equivalence, we first study Eqs.~\eqref{eq:FokkerPlanck}-\eqref{eq:diffusion_final} in the single-particle limit $N=1$.
In this case, both $V^\alpha$ and $\cD^{\alpha\beta}$ exclusively depend on the auto-correlation times $\tau^{\alpha\beta}$ through:
\begin{equation}
    \cD^{\alpha\beta} = D_{\rm t} \delta^{\alpha\beta} + \frac{v^2(\bfR)}{d} \tau^{\alpha\beta} \;, \quad V^\alpha = \frac{1}{2} \partial^\beta \cD^{\alpha\beta}\;. 
    \label{eq:singleparticle-diffdrift}
\end{equation}
The proportionality between $V^\alpha$ and the divergence of the diffusion tensor $\cD^{\alpha\beta}$ can be used to determine under which conditions the macroscopic current $J^\alpha$ vanishes. 
When this occurs, Eq.~\eqref{eq:FokkerPlanck} is solved by a steady-state Boltzmann-like distribution $p_{\rm s}(\bfR) = \exp[- U(\bfR)]$ for some effective potential $U$, yielding a macroscopic equilibrium regime. 
If such a potential exists, it can be shown to satisfy $\partial^\alpha U = \frac{1}{2} (\cD^{-1})^{\alpha\gamma} \partial^\beta \cD^{\gamma\beta}$. 
We then rely on the Schwarz theorem to determine the conditions of existence of $U$, requiring $\partial^\alpha \partial^\beta U = \partial^\beta \partial^\alpha U$~\cite{o2020lamellar,o2022time,dinelli2023nonreciprocity,o2024geometric,o2025geometric,duan2025phase}. 
This eventually yields~\cite{dinelli2026PRE}:
\begin{equation}
      \mathcal{Q}^{\alpha\beta} = \mathcal{Q}^{\beta\alpha} \quad \text{where} \quad \mathcal{Q}^{\alpha\beta} := \partial^\alpha \left[ (\mathcal{D}^{-1})^{\beta\gamma} \, \partial^\delta \mathcal{D}^{\gamma\delta} \right]\;
\label{eq:generalized_schwarz}
\end{equation}
for any pair of indices $\alpha \neq \beta$. Equation~\eqref{eq:generalized_schwarz} can be interpreted as a condition on the microscopic persistence times $\tau^{\alpha\beta}$, via Eq.~\eqref{eq:singleparticle-diffdrift}, to ensure the existence of an effective equilibrium regime.

First, we note that Eq.~\eqref{eq:generalized_schwarz} is \textit{always} satisfied in the athermal case $D_{\rm t}=0$, yielding $p_{\rm s} \propto v^{-1}$: in the absence of translational noise, non-interacting particles always accumulate in regions of space where they move slower~\cite{schnitzer1993theory,tailleur2008statistical, arlt2018painting,frangipane2018dynamic,metzger2024revisiting,metzger2025exceptions}. 
For $D_{\rm t} > 0$, 
the system does not necessarily admit a large-scale equilibrium regime. 
However, if the $\Theta$-dynamics features isotropic and achiral auto-correlations, i.e. $\mathbb{C}^{\alpha\beta}(t) = C(t) \delta^{\alpha\beta}$, then 
the resulting diffusion tensor is also isotropic: $\cD^{\alpha\beta} = \cD_0 \delta^{\alpha\beta}$.
In this case, Eq.~\eqref{eq:generalized_schwarz} is satisfied, yielding an effective-equilibrium description with: 
\begin{equation}
    p_{\rm s}(\bfR) \propto \exp\biggl[- \frac{1}{2}\log \cD_0\biggr] \propto \biggl[D_{\rm t} + \frac{\tau_{\rm p}}{d}v(\bfR)^2\biggr]^{-1/2}\;,
    \label{eq:ps_Dt}
\end{equation}
where the details of the $\Theta$-dynamics only enter through the persistence time $\tau_{\rm p}$. 

\begin{center}
\begin{figure}
  \includegraphics[width=\columnwidth]{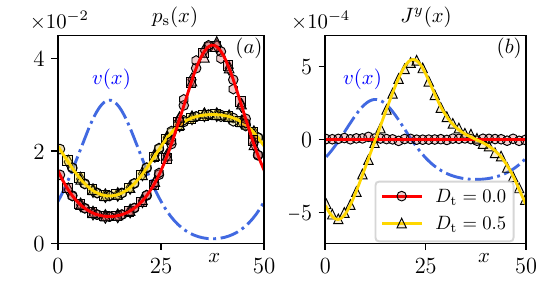}
    \caption{Steady-state particle distribution $p_{\rm s}(x)$ and current $J^y(x)$, for both $D_{\rm t} = 0$ (red) and $D_{\rm t} > 0$ (yellow). 
    Symbols represent numerical measurements from $2d$ simulations; solid lines represent theoretical predictions. ({\bf a}) Measurements of $p_{\rm s}(x)$ in a motility landscape $v(x)$. Data points correspond to standard ABPs (circle), ABPs with angular inertia (diamond), ABPs with motility-state switching (square), ABPs with fractional superdiffusive (hexagon) and subdiffusive (triangle) angular dynamics. Simulation parameters are chosen such that $\tau_{\rm p}=1$. ({\bf b}) Steady-state current $J^y(x)$ for chiral ABPs in a motility landscape $v(x)$. In both panels, $v(x) = v_0 \exp[A \sin( 2\pi x/L_x)]$ is sketched with a dash-dotted line, scaled and shifted for visualization purposes, and $v_0 = A = 1$, $L_x=50$. See the End Matter and Refs.~\cite{mandelbrot1968fractional,sprenger2023dynamics,gomez2020active,dinelli2026PRE} for further details.} 
    \label{fig:single-particle}
\end{figure}
\end{center}

To test our results, we perform $2d$ particle-based simulations of active particles in an activity landscape $v(x)$ that depends on a single spatial coordinate. 
\if{We consider a variety of examples for the orientational dynamics \plm{such as:...go to list and drop sentence in the middle}, including cases whose coarse-grained description had not previously been derived in the literature. 
More specifically, we measure $p_{\rm s}(x)$ for {(i)} standard active Brownian particles (ABPs), {(ii)} ABPs with angular inertia, {(iii)} ABPs switching between $K$ states with different internal persistence times, {(iv-v)} ABPs whose orientations undergo super- and sub-diffusive fractional Brownian motion. Details on the dynamics can be found in the End Matter and~\cite{mandelbrot1968fractional,sprenger2023dynamics,caprini2022role,gomez2020active,dinelli2026PRE}. 
In Fig.~\ref{fig:single-particle}a we show how all steady-state distributions $p_{\rm s}(x)$ collapse onto a single master curve, predicted via Eq.~\eqref{eq:ps_Dt}, when the underlying microscopic processes share the same $\tau_{\rm p}$.
\fi
We consider a variety of examples for the $\Theta$-dynamics, detailed in the caption of Fig.~\ref{fig:single-particle} and in the End Matter. In Fig.~\ref{fig:single-particle}a, we show that the corresponding distributions $p_{\rm s}(x)$ collapse onto a single master curve, predicted via Eq.~\eqref{eq:ps_Dt}, when the underlying microscopic processes share the same $\tau_{\rm p}$. 
%
%
Notably, we find an excellent agreement also for non-Markovian processes with finite $\tau_{\rm p}$, such as for fractional Brownian dynamics of $\Theta$~\cite{mandelbrot1968fractional}.

Finally, our coarse-grained theory quantitatively captures steady-state currents for systems that violate the equilibrium condition of Eq.~\eqref{eq:generalized_schwarz}. 
To show this, we consider a system of chiral ABPs in $2d$ subjected to an activity profile $v(x)$. 
Due to chirality, the corresponding auto-correlation tensor $\mathbb{C}^{\alpha\beta}(t)$ exhibits anti-symmetric contributions: $\mathbb{C}^{xy}(t) = -\mathbb{C}^{yx}(t)$~\cite{hargus2021odd,kalz2024field}. 
Eqs.~\eqref{eq:FokkerPlanck}-\eqref{eq:diffusion_final} can be used to compute the steady-state current $J^y(x)$ and density $p_{\rm s}(x)$, see End Matter and~\cite{dinelli2026PRE}.
Direct comparison with microscopic simulations, reported in Fig.~\ref{fig:single-particle}b, shows an excellent match both for zero and finite $D_{\rm t}$. 


\noindent\emph{Large-scale equlibrium for active polymers.}---We now turn to the case of active polymers, i.e., $N>1$.
For simplicity, we focus on systems with isotropic and achiral $\Theta$-dynamics, so that $\mathbb{C}^{\alpha\beta}_{ij}(t) = C_{ij}(t) \delta^{\alpha\beta}$, and defer the more general discussion to our companion paper~\cite{dinelli2026PRE}.
Hence, the diffusion tensor of Eq.~\eqref{eq:diffusion_final} is also isotropic, i.e., $\cD^{\alpha\beta} = \cD_0 \delta^{\alpha\beta}$.
As a consequence, the drift given by Eq.~\eqref{eq:drift_final} is related to the diffusivity gradient as:
\begin{equation}
    V^{\alpha} = \partial^\alpha \cD_0\left( 1 - \epsilon/2 \right)\;,
    \label{eq:eps_relation}
\end{equation}
where:
\begin{equation}
    \epsilon = 2 - \dfrac{N \sum_{ijk} \varphi_{ji} \varphi_{jk} \int_0^\infty \rmd t e^{-\lambda_j t} C_{ij}(t)}{\sum_{ij} \int_0^\infty \rmd t \> C_{ij}(t)} \;.
    \label{eq:eps_value}
\end{equation}
Under these conditions, it can be shown that the system always admits a macroscopic effective equilibrium regime featuring zero steady-state current and the following Boltzmann-like distribution for the center of mass:
\begin{equation}
    p_{\rm s}(\bfR) \propto \exp \biggl[ - \frac{\epsilon}{2} \log \cD_0 \biggr] \propto [\cD_0(\bfR)]^{-\epsilon/2} \;.
    \label{eq:ps_Polymer}
\end{equation}
Equation~\eqref{eq:ps_Polymer} reveals an interesting behavior of active polymers, namely that they accumulate in high-activity regions (maxima of $\cD_0$) when $\epsilon < 0$, in stark contrast with single particles~\cite{vuijk2021chemotaxis,muzzeddu2024migration, ravichandir2025transport}.
Indeed, as can be seen from  Eq.~\eqref{eq:singleparticle-diffdrift}, the latter are characterized by $\epsilon = 1$.
Moreover, Eq.~\eqref{eq:eps_value} shows that $\epsilon$ is controlled by the polymer structure ($N$, $\varphi_{ij}$), interaction stiffness ($\lambda_j$), and orientational correlations $C_{ij}(t)$, thereby providing multiple routes to invert accumulation from low to high activity.
This mechanism directly sets the stage for the emergence of new collective phenomena in interacting systems, as we now demonstrate.
\begin{center}
\begin{figure}
  \includegraphics[width=\columnwidth]{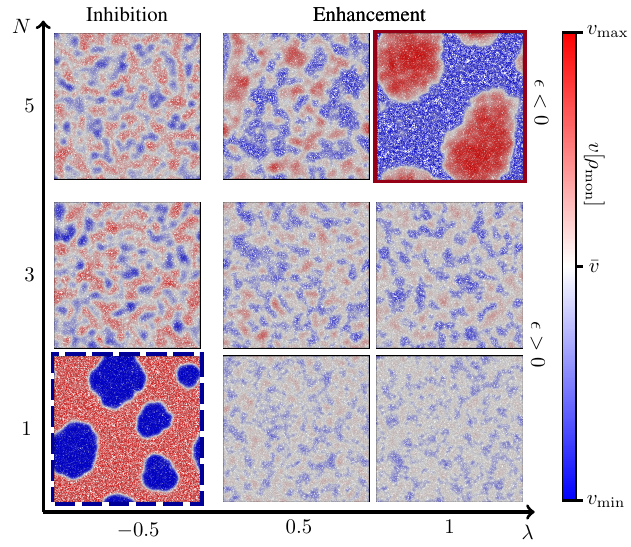}
    \caption{Emergence of MIPS and anti-MIPS in quorum-sensing active polymer chains, consisting of $N$ ABP monomers. Snapshots of the microscopic dynamics show anti-MIPS for large $N$ and motility enhancement (solid box, top right), where the dense phase is more active than the dilute one. Conventional MIPS is recovered at small $N$ under motility inhibition (dashed box, bottom left). Particles are colored by their local speed $v[\rho_{\rm mon}]$, ranging from $v_{\rm min} = v_0 e^{-|\lambda|}$ to $v_{\rm max} = v_0 e^{|\lambda|}$; in a homogeneous system with monomer density $\rho_{\rm mon,0}$, the speed is $\bar v$. Parameters: $L_x = L_y = 24$, $D_r = \gamma = \kappa = 1$, $v_0 = 1$. Further numerical details are given in End Matter.} 
    \label{fig:antiMIPS}
\end{figure}
\end{center}
\vspace{-0.8cm}
\noindent\emph{Anti-MIPS in quorum-sensing active polymers.}---To illustrate the implications of high-activity accumulation, we conclude our study by extending our coarse-grained descriptions to active polymers interacting via quorum sensing (QS), by which active particles adapt their motility based on the local density of their peers~\cite{miller2001quorum}. 
%
%
QS provides a paradigmatic example of motility regulation, and is known to trigger motility-induced phase separation (MIPS)~\cite{tailleur2008statistical,cates2015motility,bauerle2018self,curatolo2020cooperative,duan2023dynamical,lefranc2025synthetic,burekovic2026active}, making it an ideal framework to study collective phenomena in active polymers.

We consider a system of $M$ active polymers, each composed of $N$ monomers with positions $\bfr_{i,n}$ and orientations $\bfu_{i,n}$, where $i$ indicates the monomer index within polymer $n$. 
We consider isotropic and achiral orientational dynamics with no inter-polymer correlations. 
In the presence of QS interactions, each monomer adapts its self-propulsion speed $v$ to the local monomer density $\rho_{\rm mon}(\bfr)=\sum_{i,n}\delta(\bfr-\bfr_{i,n})$. 
The dynamics of $\bfr_{i,n}$ is still given by Eq.~\eqref{eq:dynamics-position}, where $v$ depends on the density through:
\begin{equation}
    v(\bfr_{i,n},[\rho_{\rm mon}]) = v_0 \exp[\lambda \cS(\tilde\rho_{\rm mon}(\bfr_{i,n}))] \;.
\end{equation}
Here, $\cS(\cdot)$ is a sigmoidal increasing function, and $\tilde\rho_{\rm mon} = K \ast \rho_{\rm mon}$ denotes the convolution of the monomer density with an isotropic bell-shaped kernel $K$ of radius $\ell_{\rm int}$ (see End Matter). 
Finally, the parameter $\lambda$ determines whether each particle enhances their speed when the local density increases ($v'(\rho) > 0$ for $\lambda > 0$), or, conversely, undergoes motility inhibition $v'(\rho)<0$ for $\lambda < 0$.

To study the large-scale behavior of our system, we perform particle-based simulations for chains of ABPs, where we vary the polymer length $N$ and the QS strength $\lambda$. 
For sufficiently large $N$, we report in Fig.~\ref{fig:antiMIPS} a phase-separated regime also in the presence of motility \textit{enhancement} ($v'(\rho)>0$, $\lambda>0$). 
The corresponding dense phases are thus more active than the surrounding gas, in contrast with conventional MIPS, where only motility inhibition can drive phase separation~\cite{tailleur2008statistical,fily2012athermal,cates2015motility}.
We thus denote this new phase as \emph{anti}-motility-induced phase separation or anti-MIPS.

To elucidate its origin, we apply our coarse-graining procedure to obtain the stochastic hydrodynamics of the system to lowest order in gradients.
As we detail in~\cite{dinelli2026PRE}, we supplement our perturbative approach with standard techniques~\cite{dinelli2024fluctuating} to obtain the time evolution of the fluctuating density of polymers, defined from the CM positions $\{R_n^\alpha\}$ as $\rho(\bfr) = \sum_{n=0}^{M-1} \delta(\bfr - \bfR_n)$. All in all, this reads:
\begin{equation}
    \partial_t \rho = \partial^\alpha [ M_0 \partial^\alpha \mu(\bfr,[\rho]) + \sqrt{2  M_0} \Lambda^\alpha(\bfr,t)] \;,
    \label{eq:hydroQS}
\end{equation}
where $\Lambda^\alpha(\bfr,t)$ is a Gaussian white noise field with zero mean and $\langle \Lambda^\alpha(\bfr,t)\Lambda^\beta(\bfr',t') \rangle=\delta(\bfr-\bfr') \delta(t-t') \delta^{\alpha\beta}$, and $M_0$ is a density-dependent mobility~\cite{dinelli2026PRE}. Here, the effective chemical potential $\mu$ reads:
\begin{equation}
    \mu(\bfr,[\rho]) = \epsilon \log v(\bfr, [N \rho]) + \log \rho(\bfr) \;,
    \label{eq:chempot}
\end{equation}
where $\epsilon$ is given by Eq.~\eqref{eq:eps_value}. Equations~\eqref{eq:hydroQS},~\eqref{eq:chempot} can then be used to predict the large-scale organization of our system. Indeed, under a local approximation $\log v(\bfr, [N \rho]) \simeq \log v(N \rho(\bfr))$, the hydrodynamics~\eqref{eq:hydroQS} becomes an equilibrium model-B field theory~\cite{hohenberg1977theory}. The chemical potential then derives from an effective free energy density, $\mu(\rho) = f'(\rho)$, where:
\begin{equation}
    f(\rho) = \epsilon \int^\rho \log v(N s) \rmd s + \rho \log \rho \;.
    \label{eq:free_energy_density}
\end{equation}
%
Phase separation requires that $f''(\rho)<0$ over some density range~\cite{chaikin1995principles}. 
For single particles, where $N=\epsilon=1$, this occurs only for $v'(\rho) < 0$. 
By contrast, active polymers accumulate in regions of high activity when $\epsilon < 0$, so that $f$ can become concave even for motility enhancement, $v'(\rho) > 0$. 
In this case, a local increase in $v$ promotes further polymer accumulation, resulting in a positive feedback loop that ultimately drives phase separation.

\begin{center}
\begin{figure}
  \includegraphics[width=\columnwidth]{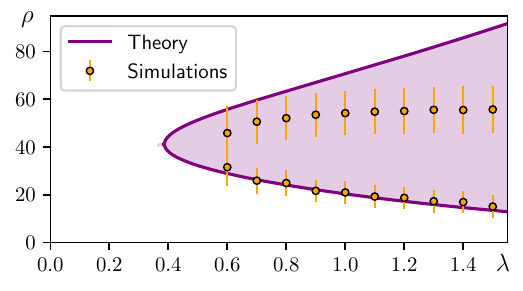}
    \caption{Phase diagram for anti-MIPS in QS active-polymer chains of size $N=5$ in the $(\lambda,\rho)$-space. Symbols correspond to numerical measurements of the binodals, while shaded region is obtained from the common-tangent construction on the effective free energy $f(\rho)$ using Eqs.~\eqref{eq:eps_value} and \eqref{eq:free_energy_density}. Simulation parameters: $L_x = L_y = 30$, $D_r = \gamma = \kappa = 1$, $v_0 = 1$. Further numerical details are given in End Matter.} 
    \label{fig:phasediagram}
\end{figure}
\end{center}

To test this idea, we consider a case where $\epsilon < 0$ and build the phase diagram using the equilibrium common-tangent construction on $f(\rho)$. 
In Fig.~\ref{fig:phasediagram} we compare the coexisting densities, measured in simulations of pentamers ($N=5$) with motility enhancement, to the predicted phase diagram. 
Our theory indeed accurately captures the low density phase, while it fails at quantitively accounting for the liquid density. 
This is expected as we neglect higher-order gradients in the hydrodynamics~\cite{solon2018generalized,burekovic2026active}.
\vspace{-0.9cm}
\emph{Discussion.}--- In this work, we derived a unified hydrodynamics for scalar active matter with motility regulation, revealing a large-scale equivalence across a broad class of self-propelled systems.
Our results show that controlling microscopic orientational correlations offers a general strategy to engineer collective behavior in active materials,  
with potential applications in the design of soft matter with motility regulation, from light-controlled bacteria~\cite{di2010bacterial,arlt2018painting,frangipane2018dynamic,pellicciotta2023light} to quorum-sensing colloidal systems~\cite{bauerle2018self,lefranc2025synthetic}.
While we focused here on spatially modulated activity, the same perturbative approach naturally extends to other forms of motility regulation, including taxis~\cite{dinelli2026PRE}.

Building on our results, we revealed a new form of motility-induced phase separation, or anti-MIPS, in QS polymeric systems. 
The resulting hydrodynamic description is currently limited to lowest order in gradients. However, systematic frameworks exist to derive hydrodynamic theories beyond leading-order gradients~\cite{cates2015motility,solon2018generalized,duan2023dynamical}, including recently-developed perturbative approaches~\cite{burekovic2026active,kafri2026externalpotential}. We thus expect a higher-order theory for anti-MIPS to be also within reach in future research.

Finally, our approach relies on a dilute regime where steric repulsion is negligible. It would thus be interesting to study how anti-MIPS and repulsive forces interplay with one another. In this regime, nematic order may also emerge, as observed in related systems~\cite{abaurrea2018collective,dedenon2026importance}. This problem could be approached, for instance, relying on recent methods available in the literature~\cite{nguyen2025contact}.

\emph{Acknowledgments.---}The authors thank Karsten Kruse for insightful discussions and support. This work was partially funded by Swiss National Science Foundation through grant number $200020\mathrm{E}\_219164$.

\bibliography{biblio}

\newpage
\section{End matter}
\emph{Single-particle simulations}.—
To produce Fig.~\ref{fig:single-particle}, we integrate Eq.~\eqref{eq:dynamics-position} for $N=1$ in $2d$ using an Euler scheme, in a box of size $L_x \times L_y = 50 \times 10$ with periodic boundary conditions. Unless specified otherwise, we simulate $5\times 10^3$ trajectories up to time $T=10^6$ with time step $dt=0.005$ and $\gamma=1$. The self-propulsion speed is modulated as $v(x)= v_0\exp[A \sin( 2\pi x / L_x)]$ with $v_0 = 1$, $A = 1$, and orientations are identified by the unit vector $\bfu=(\cos \theta, \sin \theta)$.

For Fig.~\ref{fig:single-particle}a, we consider five distinct angular dynamics, choosing parameters such that the persistence time $\tau_{\rm p}=1$:

\noindent (i) Standard ABP:
\begin{equation*}
    \dot\theta = \sqrt{2 D_r} \xi(t)\,,
\end{equation*}
with $\tau_{\rm p} = D_r^{-1}$ and $D_r=1$.

\smallskip

\noindent (ii) ABP with angular inertia:
\begin{eqnarray*}
    \dot\theta = \omega(t)\;, \qquad I\dot\omega = -\omega + \sqrt{2 D_r} \xi(t)\,.
\end{eqnarray*}
The persistence time reads $\tau_{\rm p} =  I e^{z} z^{-z} \Gamma(z,0,z)$ with $z=I D_r$~\cite{sprenger2023dynamics,caprini2022role}. We set $D_r = 2$, $I = 0.7789084214$.

\smallskip

\noindent (iii) ABP switching between $K$ motility states. Each state $\mu$ is identified by an internal persistence time $\tau_\mu$. Arrested states ($\tau_{\mu} = 0$) correspond to $\bfu=0$, whereas, in active states with $\tau_\mu > 0$: 
    \begin{equation*}
        \dot\theta = \sqrt{2 \tau_{\mu}^{-1}} \> \xi(t) \;.
    \end{equation*}
Switching events between any two states $\mu \to \nu$ occur at a Poisson rate $\kappa_{\mu \to \nu}$. 
Upon switching from an arrested to an active state, the new orientation $\theta$ is randomly sampled in $[0, 2\pi)$; any other type of transition $\tau_\mu \to \tau_\nu$ leaves the orientation unaltered. 
In our simulations, we consider a $3$-state cycle where the only non-zero rates  are $\kappa_{0\to1}=\kappa_{1\to2}=\kappa_{2\to0}=k$. The overall persistence time yields~\cite{dinelli2026PRE}:
\begin{equation*}
    \tau_{\rm p} = \frac{(\tau_0+\tau_1 + \tau_2) + 3k(\tau_0\tau_1 + \tau_1\tau_2 + \tau_2\tau_0) + 9 k^2 \tau_0 \tau_1 \tau_2}{3[1+k (\tau_0 +\tau_1 + \tau_2) + k^2 (\tau_0 \tau_1 + \tau_1 \tau_2 + \tau_2 \tau_0)]}\,.
\end{equation*}
We use $k=0.5$, $\tau_0 = 4$, $\tau_1 = 0$, $\tau_2 = 2$.

\smallskip

\noindent (iv)-(v) Fractional Brownian angular noise~\cite{mandelbrot1968fractional,gomez2020active}:
\begin{equation*}
    \dot\theta = \sqrt{2D_H} \xi(t)\,,
\end{equation*}
with covariance
\begin{equation*}
    \langle \xi(t) \xi(s)\rangle = H |t-s|^{2H-1} \left[\frac{2H-1}{|t-s|}+2\delta(t-s) \right]\,.
\end{equation*}
The Hurst parameter $H$ takes values in $(0,1)$. The case $H=1/2$ corresponds to Brownian motion for $\theta$. 
The persistence time is $\tau_{\rm p} = z \Gamma(z) D_H^{-z}$ with $z=1/(2H)$~\cite{gomez2020active}. We simulate the dynamics in a subdiffusive ($H=0.3$, $D_H=1.2777573545$) and a superdiffusive case ($H=0.7$, $D_H=0.8782298700$). The full noise trajectories are generated prior to the simulation using the Python package \texttt{fbm} over $T=2\times 10^4$ with $dt=0.005$.

\medskip

For Fig.~\ref{fig:single-particle}b, we simulate chiral ABPs:
\begin{equation*}
    \dot\theta = \omega + \sqrt{2 D_r} \xi(t)\,.
\end{equation*}
Defining $\tau^{||} = {D_r}/(D_r^2 + \omega^2)$ and $\tau^{\rm a} = {\omega}/(D_r^2 + \omega^2)$, our theory predicts~\cite{dinelli2026PRE} $p_{\rm s}(x) \propto  \left( D_{\rm t} + \frac{v^2(x)}{d} \tau^{||} \right)^{-1/2}$ and
\begin{equation*}
    J^x = 0\,, \quad 
    J^y(x) = -\frac{\tau^{\rm a}}{4} p_{\rm s}(x) \partial_x v^2 \left[ 1 - \frac{\tau^{||} v^2(x)}{2 D_{\rm t} + \tau^{||} v^2(x)}\right]\,.
\end{equation*}
We set $\omega = D_r = 1$. Currents are measured by binning the system into stripes of width $\delta x$ and averaging particle velocities along over time.

\medskip

\emph{Simulations of QS polymers}.—
We integrate Eq.~\eqref{eq:dynamics-position} for ABP chains of length $N$ in $2d$ , using an Euler scheme. We take periodic boundary conditions and adaptive time step $dt = \min \big[ dt_{\rm max}, \ell_{\rm int}/(10 v_{\rm max}) \big]$, with $dt_{\rm max}=10^{-3}$. At each iteration and for each particle $(i,n)$, the self-propulsion speed $v$ is updated based on the local monomer density $\tilde\rho_{\rm mon}(\bfr_{i,n})$ through:
\begin{equation*}
    v(\tilde\rho) = v_0 \exp\bigl[ \lambda \tanh\bigl( \frac{\tilde{\rho}-\rho_{\rm t}}{\varphi} \bigr)\bigr]\,.
\end{equation*}
The monomer density around each particle $(i,n)$ is computed as $\tilde\rho_{\rm mon}(\bfr_{i,n}) = \sum_{(j,m) \neq (i,n)} K(\bfr_{i,n}-\bfr_{i,m})$, using spatial hashing. The coarse-graining kernel $K$ is $K(r)=\frac{1}{Z}\exp\left(\frac{\ell_{\rm int}^2}{\ell_{\rm int}^2-r^2}\right)\Theta(\ell_{\rm int}-r)$ with $\ell_{\rm int}=1$ and $Z$ a normalization constant such that $\int d \bfr K(\bfr)=1$. We set $v_0 = 1$, $\rho_{\rm t} = 200$, $\varphi = 50$, $\gamma=\kappa=1$, $D_r = 1$.  

Simulations in Fig.~\ref{fig:antiMIPS} are initialized with a homogeneous condition with initial monomer density $\rho_{\rm mon, 0} = 180$. Simulations in Fig.~\ref{fig:phasediagram} are initialized as a phase-separated band of liquid immersed in a surrounding gas; the fraction of liquid phase is set to $0.3$ of the total volume, and, for each $\lambda$, the initial densities of liquid and gas are set from the theoretical phase diagram.

\emph{Expression of $\epsilon$ for an ABP chain.}—Using Eq.~\eqref{eq:eps_value} we compute the parameter $\epsilon$ for a linear chain of length $N$, see also~\cite{muzzeddu2024migration,ravichandir2025transport}. We obtain:
\begin{equation}
    \epsilon = 1 - \sum_{n=1}^{N-1} \frac{1}{1 + 4 \frac{\kappa D_r}{\gamma} \sin^2[n \pi/(2 N)]} \;,
\end{equation}
which is a decreasing function of $N$. For our choice of parameters in Figs.~\ref{fig:antiMIPS}-\ref{fig:phasediagram}, $\epsilon < 0$ for $N \geq 4$.

\end{document}